\documentclass[fleqn,twoside]{article}
\usepackage{espcrc2}
\usepackage{epsfig}

\title{\hfill \parbox{6em}{\normalsize DESY-03-077 \\ 
		hep-ph/0306311 \\ \\} \\ 
Exclusive two-photon annihilation at large energy or large
virtuality\thanks{Presented at the International Conference on the
Structure and Interactions of the Photon (PHOTON 2003), Frascati,
Italy, 7--11 April 2003.  To appear in the proceedings.}}

\author{M. Diehl \address{Deutsches Elektronen-Synchroton DESY,
        22603 Hamburg, Germany}%
}
       
\begin{document}

\begin{abstract}
I review recent progress in the theory of $\gamma\gamma$ annihilation
into meson or baryon pairs at large energy, and of the process
$\gamma^*\gamma^* \to \pi^0$ at large photon virtuality.
\vspace{1pc}
\end{abstract}

\maketitle



\section{INTRODUCTION}

In this contribution I discuss two topics of two-photon physics.  In
both of them, handbag diagrams play a major role, but the physics
issues are rather different.

Several talks in this session have shown the ongoing experimental
progress in measuring $\gamma \gamma$ annihilation into meson or
baryon pairs at large energy.  Recent theoretical work suggests that
these processes might help clarify the long-standing problems in
understanding the dynamics of exclusive reactions in fixed-angle
kinematics (i.e.\ at large Mandelstam variables $s$, $t$, $u$).  In a
second part I investigate what one might learn from the annihilation
of two virtual photons into a single $\pi^0$, beyond what we have
already learned from the measurements with one virtual photon and
their theoretical analysis.

This presentation is based on work with P.~Kroll and C.~Vogt
\cite{Diehl:2001fv,Diehl:2002yh,Diehl:2001dg}, where further detail
and references can be found.


\section{REAL PHOTONS AND LARGE $s$, $t$, $u$}
\label{sec:real}

\begin{figure}[htb]
\begin{center}
	\leavevmode
	\epsfxsize=0.43\textwidth
	\epsfbox{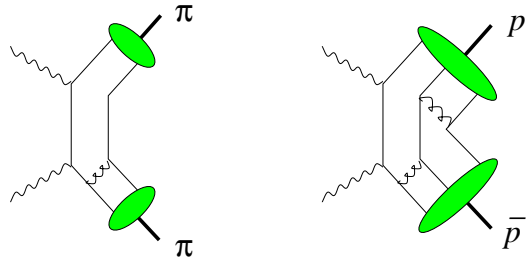}
\end{center}
\caption{Example graphs for $\gamma\gamma \to \pi\pi$ and for
$\gamma\gamma \to p\bar{p}$ in the hard-scattering mechanism.}
\label{fig:hard}
\end{figure}

Exclusive two-photon processes like $\gamma\gamma \to \pi\pi$ or
$\gamma\gamma \to p\bar{p}$ are described by the hard-scattering
mechanism of Brodsky and Lepage in the limit $s\to \infty$ at fixed
$t/s$ and $u/s$ \cite{Brodsky:1981rp}.  In this limit the process
amplitudes factorize into a subprocess involving quarks and gluons and
into the distribution amplitudes for the lowest Fock states of the
produced hadrons.  Example graphs are shown in Fig.~\ref{fig:hard}.

In evaluating these graphs, one encounters regions of loop momentum
where all parton lines except those attached to one or both photons
are soft, with momenta of a few 100~MeV in the collision c.m.  In
these regions the assumptions and approximations of the
hard-scattering calculation break down and the result is not
trustworthy.  In the limit specified above this is not a problem: the
contribution from such soft regions is power suppressed in $1/s$ and
thus harmless within the accuracy of the calculation.  Problems do
however occur in practical calculations for $s$ in the range of a few
to a few 10 GeV$^2$.  Depending on what one takes for the distribution
amplitudes of the produced particles, the result either undershoots
data by an order of magnitude or more, or it receives substantial
contributions from the soft phase space regions where the result
cannot be trusted.

\begin{figure}[htb]
\begin{center}
	\leavevmode
	\epsfxsize=0.2\textwidth
	\epsfbox{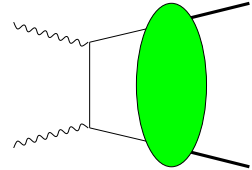}
\end{center}
\caption{Handbag graph for the soft annihilation contribution to
$\gamma\gamma \to \pi\pi$ or $\gamma\gamma \to p\bar{p}$.}
\label{fig:soft}
\end{figure}

One may take this as an indication that the soft contribution just
described plays an important role at experimentally accessible values
of $s$, and there are recent theoretical attempts to calculate this
contribution in a consistent way
\cite{Diehl:2001fv,Diehl:2002yh,Freund:2002cq}.  To justify the
approximations of the calculations we must still require that $s$,
$t$, $u$ be large compared with a typical hadronic scale; in
particular one has to avoid the region where $\sqrt{s}$ is close to
resonance masses and the regions of forward or backward scattering
angle $\theta$ in the c.m.  The separation of the dynamics into soft
and hard pieces now differs from the one in Fig.~\ref{fig:hard} and is
shown in Fig.~\ref{fig:soft}: a hard subprocess $\gamma\gamma \to
q\bar{q}$ is followed by soft hadronization into the final state.  In
order for the second step to be soft, the initial quark and antiquark
must each carry approximately the full four-momentum of one
final-state hadron; all other partons involved in the hadronization
process must have soft four-momenta in the c.m.  This soft subprocess
can be described by matrix elements of the form $\langle \pi\pi |\,
\bar{q}_\alpha(0)\, q_\beta(z) | 0\rangle$ between the vacuum and the
final state, with a quark-antiquark operator separated along the
light-cone $z^2=0$.  Such matrix elements are parameterized by
generalized distribution amplitudes \cite{Muller:1998fv,Diehl:1998dk},
which are related to generalized parton distributions by crossing
symmetry.

In the limit $s\to \infty$ the soft annihilation contribution is power
suppressed in $1/s$ compared with the hard-scattering contribution,
but it is in finite $s$ that we are interested in.  At present we have
no theory for consistently adding the contributions from the two
mechanisms without double counting.  Since in the kinematics of
interest the hard-scattering results are small (at least when the
region of soft loop momenta is discarded) I will in the following
neglect this contribution.  I will in fact assume that the soft
annihilation mechanism dominates over all other mechanisms (for
instance of vector dominance type, where the hadronic components of
the photons take part in the reaction).  This leads to predictions
which can be confronted with data.

\subsection{Results: meson pairs}

For pseudoscalar mesons $P$ the soft annihilation mechanism gives the
$\gamma\gamma \to P\bar{P}$ cross section as \cite{Diehl:2001fv}
\begin{equation}
  \label{meson-X-sect}
\frac{d\sigma(P\bar{P})}{dt} 
= \frac{8\pi \alpha_{\mathrm{em}}^2}{s^2}\,
  \frac{1}{\sin^4\theta}\, | R_{2P}(s) |^2 .
\end{equation}
The information on the soft subprocess is contained in $R_{2P}(s)$,
which is a form factor of the local operator
\begin{equation}
  \label{meson-ff}
\sum_q e^2_q\, \Big(\bar{q} \gamma^\mu i
\raisebox{0.09em}{$\stackrel{\raisebox{-0.03em}{$\scriptstyle
		\leftrightarrow$}}{D}$}{}^\nu
q + \{ \mu \leftrightarrow \nu \} \Big)
\end{equation}
and hence related with the quark energy-momentum tensor.  The
mechanism makes a \emph{prediction} for the cross section dependence
on the scattering angle $\theta$ in the two-photon c.m., keeping in
mind that the result (\ref{meson-X-sect}) is not valid for $\theta$
close to $0^\circ$ or $180^\circ$.

A general feature of the handbag graphs (independently of the
approximations needed to evaluate them) is that the reaction proceeds
through a single $q\bar{q}$ pair in the $s$-channel.  This leads to
\emph{predictions} of the soft annihilation mechanism for the
production ratios of different meson pairs.  The cleanest of them
makes only use of isospin symmetry and is
\begin{equation}
  \label{pion-ratio}
\frac{d\sigma(\pi^0\pi^0)}{dt} = \frac{d\sigma(\pi^+\pi^-)}{dt} .
\end{equation}
This result is easily derived: the two-photon collision produces the
pion pair only in a $C$-even state, which must have isospin $I=0$ or
$I=2$.  The $I=2$ state is however forbidden in the handbag mechanism
since it is unaccessible for a $q\bar{q}$ pair.  The prediction
(\ref{pion-ratio}) stands out against the hard-scattering mechanism,
where one obtains $\sigma(\pi^0\pi^0) \ll \sigma(\pi^+\pi^-)$ for pion
distribution amplitudes close to the asymptotic one
\cite{Brodsky:1981rp}.  For $\rho\rho$ production we correspondingly
have
\begin{equation}
  \label{rho-ratio}
\frac{d\sigma(\rho^0\rho^0)}{dt} = \frac{d\sigma(\rho^+\rho^-)}{dt} .
\end{equation}
Since the isospin argument works at the amplitude level, the relation
(\ref{rho-ratio}) readily generalizes to the cross sections for
definite polarization states of the $\rho$ mesons.  To my knowledge, a
prediction of the hard-scattering mechanism is unfortunately not
available for the $\rho\rho$ channel.  Returning to pseudoscalars, one
obtains as a further consequence of the handbag mechanism that
\begin{equation}
  \label{kaon-ratio}
\frac{d\sigma(K^0 \bar{K}^0)}{dt} \simeq 
	\frac{4}{25}\, \frac{d\sigma(K^+ K^-)}{dt} ,
\end{equation}
where the $\simeq$ sign signals that to obtain this result one needs
SU(3) flavor symmetry, which is only approximately satisfied.  A
further consequence of SU(3) symmetry is
\begin{equation}
\frac{d\sigma(K^+ K^-)}{dt} \simeq \frac{d\sigma(\pi^+\pi^-)}{dt} ,
\end{equation}
which---contrary to the previous relations---is independent of the
reaction mechanism.  Its violation in the real world may be taken as a
measure of how strongly SU(3) flavor symmetry is broken in this type
of process.  Notice that the often cited ratio $\sigma(K^+ K^-)
/\sigma(\pi^+\pi^-) = (f_K /f_\pi)^4$ is \emph{not} a prediction of
the hard-scattering mechanism, but assumes in addition that the
distribution amplitudes of pions and kaons have the same shape---an
assumption one may or may not wish to make.

We cannot presently calculate the form factors $R_{2P}(s)$ within QCD,
so that we have \emph{no prediction} for the $s$ dependence of the
cross section, nor for its absolute size.  We found that the
preliminary data for $\pi^+\pi^-$ production from ALEPH
\cite{Photon:2001al} and DELPHI \cite{Photon:2001de} can be described
by a form factor $R_{2\pi}(s)$ which for $s$ between 6 and 30~GeV$^2$
behaves as $s^{-1}$ and is similar in size to the electromagnetic pion
form factor $F_{\pi}(s)$ at large timelike $s$.  There is no reason
for these form factors to be equal, since they belong to different
currents, but it seems plausible that they should be of similar size.
Notice that both $R_{2\pi}(s)$ and $\sigma(\pi^+\pi^-)$ approximately
follow the $s$ dependence obtained in the hard scattering picture
(which fails badly for the absolute normalization).  That this might
happen over a finite interval in $s$ is not in contradiction with the
soft annihilation mechanism.  I note that this mechanism has an analog
in the crossed channel, namely the Feynman mechanism for spacelike
form factors and large-angle Compton scattering, see
\cite{Kroll:2003cb} for a recent overview.  In these cases, a model
study has in fact shown how the soft mechanism can systematically
mimic the hard-scattering scaling behavior in the $t$ region of
several GeV$^2$ \cite{Diehl:1999tr}.

\subsection{Results: baryon pairs}

For a baryon $B$ the soft annihilation mechanism leads to a
$\gamma\gamma \to B\bar{B}$ cross section \cite{Diehl:2002yh}
\begin{eqnarray}
  \label{baryon-X-sect}
\frac{d\sigma(B\bar{B})}{dt} 
 &=& \frac{4\pi \alpha_{\mathrm{em}}^2}{s^2}\,
\\
&& \hspace{-4em} {}\times \frac{1}{\sin^2\theta}\, 
           \Big( | R^{B}_{\mathrm{eff}}(s) |^2 
	       + \cos^2\theta\, | R^B_V(s) |^2 \Big) 
\nonumber 
\end{eqnarray}
with
\begin{equation}
| R^{B}_{\mathrm{eff}}(s) |^2 = |R^B_A + R^B_P|^2 
	+ \frac{s}{4m_B^2}\, |R^B_P|^2 .
\end{equation} 
The form factors $R^B_A$ and $R^B_P$ belong to the axial current and
$R^B_V$ belongs to the vector current of quarks, with the different
flavors weighted by their squared electric charges as in
(\ref{meson-ff}).

As in the meson case, we cannot calculate these form factors at
present and hence have \emph{no prediction} for the $s$ dependence or
the size of the cross section.  We found the CLEO and VENUS data
\cite{Artuso:1993xk} for $p\bar{p}$ production at $s$ between $6.5$
and 11~GeV$^2$ rather well described by a form factor
$R^{p}_{\mathrm{eff}}(s)$ that falls like $s^{-2}$ and is somewhat
larger in size than the measured magnetic proton form factor $G_M(s)$
at similar $s$.  We neglected $R_V(s)$ at this stage since its
prefactor in the integrated cross section is rather small in the range
$|\cos\theta \,| \le 0.6$ of the data.  In the $s$ range under study
we find again the same $s$ behavior one would obtain in the
hard-scattering mechanism, and our discussion of the meson case
equally applies here.

Our result (\ref{baryon-X-sect}) \emph{does predict} the form of the
$\theta$ dependence in terms of one unknown parameter $| R^B_V(s)
/R^B_{\mathrm{eff}}(s) |$.  We obtain more predictions if we assume
flavor SU(3) symmetry and make further approximations, spelled out in
\cite{Diehl:2002yh}.  The production ratios of all $B\bar{B}$ channels
where $B$ is in the ground state baryon octet are then expressed in
terms of a single parameter $\rho$, which describes the relative
strength of the transitions $d\bar{d} \to p\bar{p}$ and $u\bar{u} \to
p\bar{p}$.  Taking $\rho$ between $0.25$ and $0.75$ we find fair
agreement with the data of CLEO and L3 \cite{Anderson:1997ak} for
$\Sigma^0 \bar{\Sigma}^0$ and $\Lambda \bar{\Lambda}$ production, and
predict in particular that the cross section for the mixed $\Lambda
\bar{\Sigma}^0 + \Sigma^0 \bar{\Lambda}$ channel should be much lower
than~$\sigma(p\bar{p})$.

The experimental progress reported at this meeting, both for meson and
for baryon pairs, will allow one to refine the phenomenological
studies I have reported on.  The $\theta$ dependence of the cross
section and production ratios for various channels can be used to test
the soft annihilation mechanism.  If it stands up to these tests, the
data may be used to extract the various form factors which describe
the transition from $q\bar{q}$ to simple hadronic systems at a
quantitative level.

\section{VIRTUAL PHOTONS AND $s= m_\pi^2$}

\begin{figure}[htb]
\begin{center}
	\leavevmode
	\epsfxsize=0.28\textwidth
	\epsfbox{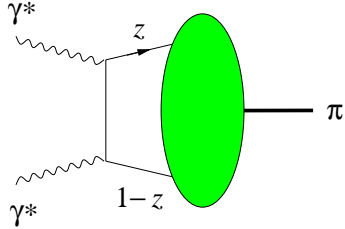}
\end{center}
\caption{Handbag graph for $\gamma^* \gamma^* \to \pi^0$ at leading
order in $\alpha_s$.}
\label{fig:pion}
\end{figure}

The annihilation of two spacelike photons into a single $\pi^0$ is one
of the simplest exclusive processes and has been studied for a long
time.  In the limit where the sum $Q^2 + Q'^2$ of the photon
virtualities becomes infinite, the amplitude factorizes into the
short-distance annihilation process $\gamma^* \gamma^*\to q\bar{q}$
and the pion distribution amplitude, as shown in Fig.~\ref{fig:pion}.
Compared to the processes discussed in Section~\ref{sec:real}, the
theory of $\gamma^*\gamma^* \to \pi^0$ is in much better shape: the
hard-scattering kernel at $O(\alpha_s)$ has been known and used to for
a long time, and the calculation of the $O(\alpha_s^2)$ corrections
has recently been reported \cite{Melic:2002ij}.  Power corrections to
the leading-twist result have been estimated by various methods and
are found to be moderate, even if one of the photon virtualities is
zero.

The pion distribution amplitude $\phi(z,\mu^2)$ is a fundamental
quantity describing the structure of the pion.  It is useful to expand
it on Gegenbauer polynomials in $2z-1$,
\begin{eqnarray}
\phi(z,\mu^2) &=& 6 z(1-z) 
\nonumber \\
&& \hspace{-4em} {}\times \Big[ 1 + 
	\sum_{n=2,4,\ldots}^\infty
	 B_n(\mu^2)\, C^{3/2}_n(2z-1) \Big] ,
\end{eqnarray}
where $z$ denotes the momentum fraction of the quark in the pion and
$\mu^2$ the factorization scale.  The cleanest experimental
constraints on the distribution amplitude come from the CLEO
measurement \cite{Gronberg:1997fj} of $\gamma^*\gamma \to \pi^0$,
where one of the photons is quasi-real.  The amplitude is
parameterized in terms of the photon-pion transition form factor,
which reads
\begin{eqnarray}
  \label{real-photon}
F_{\pi\gamma}(Q^2) &=& \frac{f_\pi}{3 \sqrt{2}\, Q^2}\, 
	\int dz\, \frac{\phi(z,\mu^2)}{z (1-z)} 
\nonumber \\
&=& \frac{\sqrt{2} f_\pi}{Q^2} 
	\Big[ 1 + \sum_{n=2,4,\ldots}^\infty B_n(\mu^2) \Big]
\end{eqnarray}
with $f_\pi \approx 131$~MeV, up to relative corrections of order
$\alpha_s$ or $1/Q^2$.  The $Q^2$ dependence of the CLEO data is well
described by this approximation already at a few GeV$^2$.  Analyzing
the data in terms of the leading-order formula (\ref{real-photon}) one
finds that the sum $\sum B_n(\mu^2)$ is small already at $\mu \sim
1$~GeV.  This conclusion does not change much when taking the
$O(\alpha_s)$ corrections into account.  Without further theoretical
assumptions one can however not conclude that the \emph{individual}
Gegenbauer moments $B_n(\mu^2)$ must be small, although this would be
in line with theoretical prejudice and with the analysis of other
processes (where however the errors of either theory or experiment are
considerably larger).  For a discussion I refer to
\cite{Diehl:2001dg}; a conflicting point of view concerning
theoretical errors is taken in \cite{Bakulev:2002uc}.  The different
$\mu$ dependence of the $B_n(\mu^2)$ provides a handle to gain
separate information about them from the $Q^2$ dependence of
$F_{\pi\gamma}$, but in order to use this one must have sufficient
control over power corrections, whose $Q^2$ dependence is much
stronger.  Data with higher statistics at larger $Q^2$ would greatly
help in this.

\begin{figure} 
\begin{center}
  \leavevmode
  \hspace{1em}
  \epsfig{file=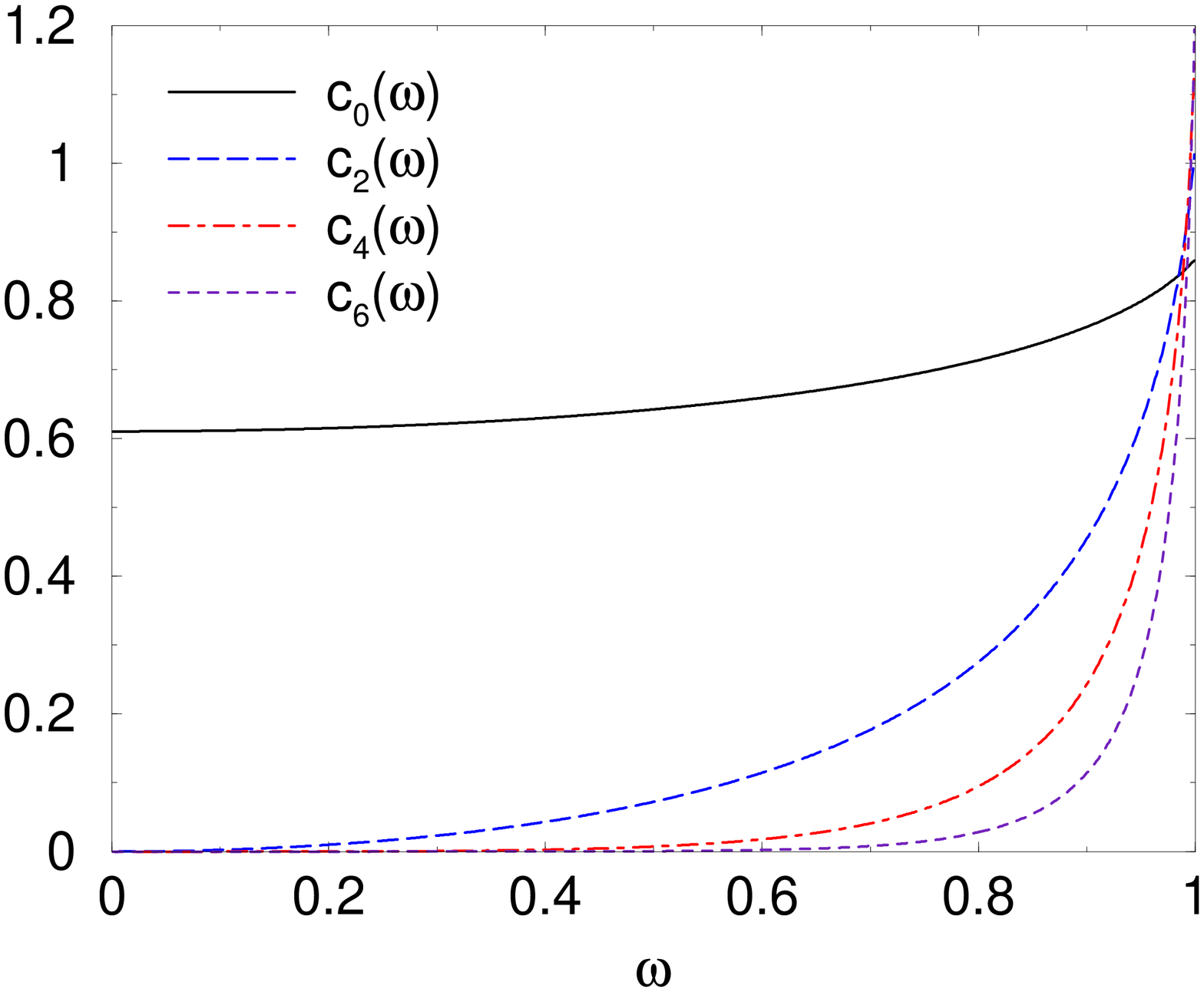,bb=90 70 560 665,width=5.4cm,angle=-90}
  \\[2em]
  \epsfig{file=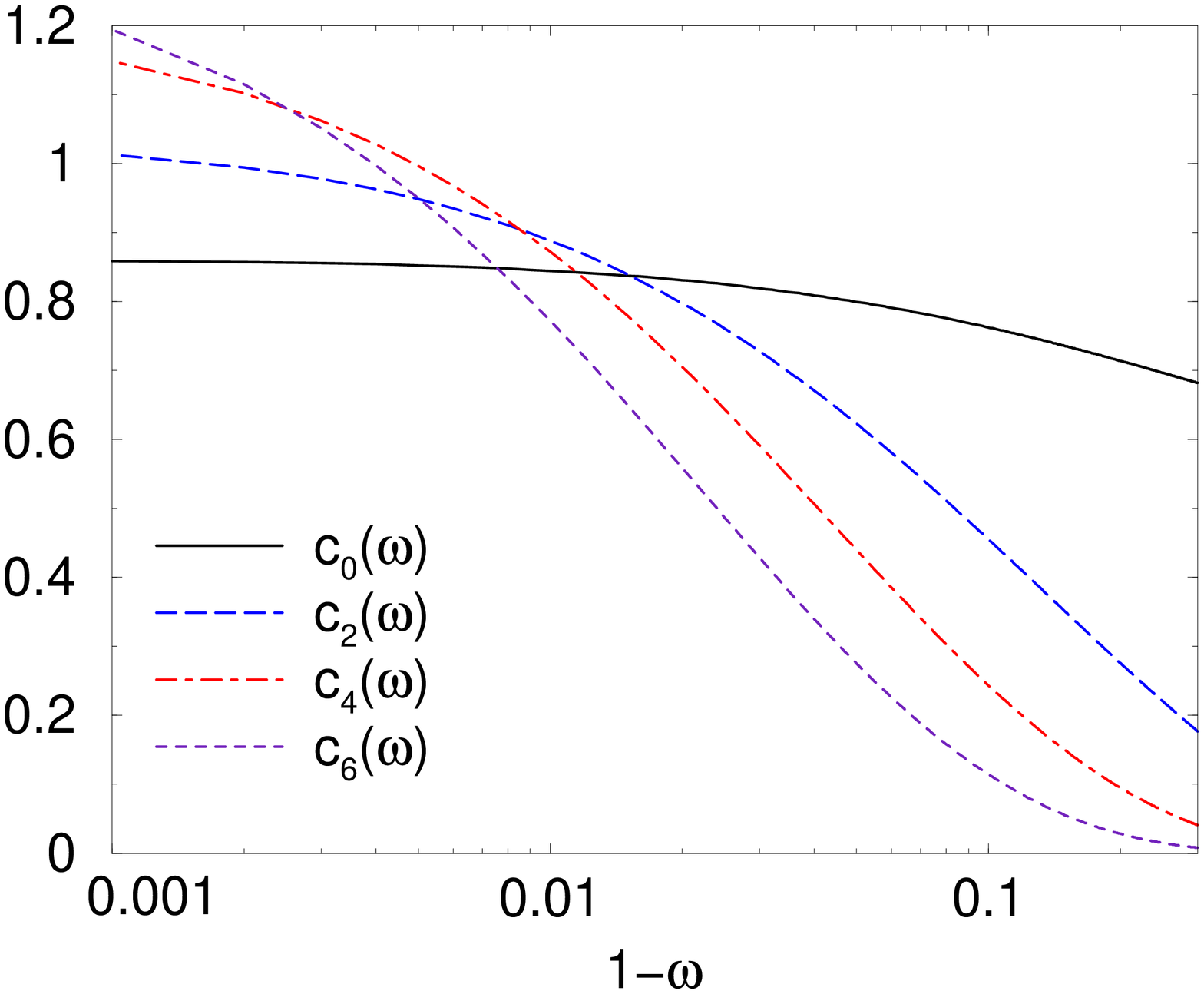,bb=90 30 560 665,width=5.4cm,angle=-90}
\end{center}
\caption{\label{fig:surprise} Coefficients $c_n$ in the expansion
(\protect\ref{virtual-photon}) of $F_{\pi\gamma^*}$.  Corrections of
$O(\alpha_s)$ are included with the factorization and renormalization
scale $\mu$ set to $\overline{Q}$ and taken as 2~GeV.}
\end{figure}

It is natural to ask whether more information can be obtained from
data on $\gamma^* \gamma^* \to \pi^0$ with both photons off-shell.  To
leading-power accuracy in $1 / \overline{Q}{}^2$, the transition form
factor for a virtual photon can be written as
\begin{eqnarray}
  \label{virtual-photon}
F_{\pi\gamma^*}(\overline{Q}{}^2,\omega) 
&=& \frac{f_\pi}{\sqrt{2}\, \overline{Q}{}^2} 
\nonumber \\
&&\hspace{-6em}  {}\times  
\Big[ c_0(\omega) + \sum_{n=2,4,\ldots}^\infty 
	 c_n(\omega,\log \overline{Q}{}^2)\, B_n \Big] ,
\end{eqnarray}
where I have suppressed the dependence on the renormalization and
factorization scale $\mu$ and chosen symmetric variables
\begin{equation}
\overline{Q}{}^2 = \frac{1}{2} (Q^2 + Q'^2) , \qquad
\omega = \frac{Q^2 - Q'^2}{Q^2 + Q'^2} .
\end{equation}
For symmetry reasons the coefficients $c_n$ are even functions of
$\omega$.  Evaluating them one finds the surprising behavior shown in
Fig.~\ref{fig:surprise}.

We see that as soon as $\omega$ becomes different from one, the
sensitivity to all but the lowest Gegenbauer moments $B_n$ rapidly
goes to zero.  For our original goal of gaining information about the
pion distribution amplitude one would thus focus on the region of
$1-\omega$ shown in the second plot of the figure.  As shown in
\cite{Diehl:2001dg}, the transition form factor in this region has
indeed some power to distinguish different scenarios for the
Gegenbauer moments which are neither implausible nor ruled out by the
existing CLEO data.

In a large region around $\omega=0$ the form factor is however hardly
sensitive to the pion structure at all, except via the pion decay
constant $f_\pi$.  In fact, one finds that it can be expanded around
$\omega=0$ as
\begin{eqnarray}
\lefteqn{
F_{\pi\gamma^*}(\overline{Q}{}^2,\omega) 
= \frac{\sqrt{2} f_\pi}{3 \overline{Q}{}^2} 
 \Bigg[ 1 - \frac{\alpha_s}{\pi} 
  + \omega^2 \Big( \frac{1}{5} 
  - \frac{1}{3} \frac{\alpha_s}{\pi} \Big)
}
\nonumber \\
&& \hspace{2em} {}+ \frac{12}{35}\, \omega^2 B_2 
   + O(\alpha_s\, \omega^2, \omega^4, \alpha_s^2) \,\Bigg]
\end{eqnarray}
where higher Gegenbauer moments $B_n$ always appear with at least a
power $\omega^n$.  Given the results of \cite{Melic:2002ij} one can
predict $F_{\pi\gamma^*}$ within QCD up to relative corrections of
order
\begin{equation}
\omega^4,\,  \alpha_s^3\, \omega^2, \, \alpha_s^4 , \, \Lambda^2 /Q^2
\end{equation}
with a hadronic scale $\Lambda$, provided one takes the lowest
coefficient $B_2$ as an input parameter.  At small enough $\omega$ one
may even neglect the terms with $\omega^2 B_2$ and then has a
prediction for $F_{\pi\gamma^*}$ only in terms of $\alpha_s$ and
$f_\pi$.  The power corrections in $\Lambda^2 /Q^2$ may at least be
estimated using the results of \cite{Khodjamirian:1997tk}, which would
require knowledge of the matrix element $\langle \pi | \bar{d}\,
g\tilde{G}^{\mu\nu} \gamma_\nu\, u | 0\rangle$.  Both this quantity
and $B_2$ are in principle amenable to calculation from first
principles in lattice QCD, and both can be constrained by
phenomenological analysis, for instance of data for the transition
form factor at $\omega$ close to 1.

The transition form factor $F_{\pi\gamma^*}$ can in this sense be
regarded as a precision observable, whose measurement would allow a
rather fundamental test of our understanding of QCD.  It is very
similar to the Bjorken sum rule for deep inelastic scattering, to
which it is intimately related as explained in \cite{Melic:2002ij}.
Note that compared with the efforts required to measure the Bjorken
sum, measurement of $F_{\pi\gamma^*}$ can in principle be done in a
single experiment for $e^+e^- \to e^+e^- \pi^0$ in suitable
kinematics.  The bad news is that the cross section is quite low:
integrating over $\omega$ from $-0.5$ to $+0.5$ one obtains a
differential $e^+e^-$ cross section of $d\sigma /d\overline{Q}{}^2
\approx 0.5$~fb~GeV$^{-2}$ at $\overline{Q}{}^2 = 4$~GeV$^2$.
Similarly low rates are obtained in the region $\omega \approx 1$
discussed above.  Experimental investigation appears therefore
difficult even at the high-luminosity machines BaBar and Belle in
their present setups.  The study of this fundamental process may
however become feasible at possible luminosity upgrades of these
facilities.



\begin{thebibliography}{99}

\bibitem{Diehl:2001fv} M.~Diehl, P.~Kroll and C.~Vogt,
Phys.\ Lett.\ B {\bf 532}, 99 (2002)
[hep-ph/0112274].

\bibitem{Diehl:2002yh}
M.~Diehl, P.~Kroll and C.~Vogt,
Eur.\ Phys.\ J.\ C {\bf 26}, 567 (2003)
[hep-ph/0206288].

\bibitem{Diehl:2001dg}
M.~Diehl, P.~Kroll and C.~Vogt,
Eur.\ Phys.\ J.\ C {\bf 22}, 439 (2001)
[hep-ph/0108220].

\bibitem{Brodsky:1981rp}
S.~J.~Brodsky and G.~P.~Lepage,
Phys.\ Rev.\ D {\bf 24}, 1808 (1981).

\bibitem{Freund:2002cq}
A.~Freund et al.,
Phys.\ Rev.\ Lett.\  {\bf 90}, 092001 (2003)
[hep-ph/0208061].

\bibitem{Muller:1998fv}
D.~M{\"u}ller et al.,
Fortsch.\ Phys.\  {\bf 42}, 101 (1994)
[hep-ph/9812448].

\bibitem{Diehl:1998dk}
M.~Diehl et al.,
Phys.\ Rev.\ Lett.\  {\bf 81}, 1782 (1998)
[hep-ph/9805380].

\bibitem{Photon:2001al}
A.~Finch, in: Proceedings of PHOTON 2001, World Scientific, Singapore
(2002).

\bibitem{Photon:2001de}
K.~Grzelak, in: Proceedings of PHOTON 2001, World Scientific, Singapore
(2002).

\bibitem{Kroll:2003cb}
P.~Kroll,
hep-ph/0302169.

\bibitem{Diehl:1999tr}
M.~Diehl et al.,
Phys.\ Lett.\ B {\bf 460}, 204 (1999)
[hep-ph/9903268].

\bibitem{Artuso:1993xk}
CLEO Collaboration,
Phys.\ Rev.\ D {\bf 50}, 5484 (1994);\\
%
VENUS Collaboration,
Phys.\ Lett.\ B {\bf 407}, 185 (1997).

\bibitem{Anderson:1997ak}
CLEO Collaboration,
Phys.\ Rev.\ D {\bf 56}, 2485 (1997)
[hep-ex/9701013];\\
%
L3 Collaboration,
Phys.\ Lett.\ B {\bf 536}, 24 (2002)
[hep-ex/0204025].


\bibitem{Melic:2002ij}
B.~Meli{\'c}, D.~M{\"u}ller and K.~Passek-Kumeri\v{c}ki,
hep-ph/0212346.

\bibitem{Gronberg:1997fj}
CLEO Collaboration,
Phys.\ Rev.\ D {\bf 57}, 33 (1998)
[hep-ex/9707031].

\bibitem{Bakulev:2002uc}
A.~P.~Bakulev et al.,
Phys.\ Rev.\ D {\bf 67}, 074012 (2003)
[hep-ph/0212250].

\bibitem{Khodjamirian:1997tk}
A.~Khodjamirian,
Eur.\ Phys.\ J.\ C {\bf 6}, 477 (1999)
[hep-ph/9712451].

\end{thebibliography}
\end{document}